\documentclass[12pt]{article}

\usepackage{amsthm,amssymb,amsmath}
\usepackage[american,british]{babel}

 1
 1

\newcommand{\bt}{{\boldsymbol{t}}}

\newcommand{\bx}{{\boldsymbol{x}}}
\newcommand{\by}{{\boldsymbol{y}}}

\newcommand{\be}{\begin{equation}}
\newcommand{\ee}{\end{equation}}
\newcommand{\gr}{\operatorname{Gr}}

\begin{document}
\title{\sc The KP Hierarchy \\ in Miwa coordinates \thanks{Partially supported by
CICYT proyecto PB95--0401 }}
\author{Boris Konopelchenko
\\ {\em Dipartimento di Fisica,
Universit\'a de Lecce}
\\ {\em 73100 Lecce, Italy}
\and
 Luis Mart\'\i nez  Alonso \\
{\em Departamento de F\'\i sica Te\'orica, Universidad
Complutense}\\{\em E28040 Madrid, Spain}}
\date{} \maketitle
\begin{abstract}
A systematic reformulation of the KP hierarchy by using continuous
Miwa variables is presented. Basic quantities and relations are
defined and determinantal expressions for Fay's identities are
obtained. It is shown that in terms of these variables the KP
hierarchy gives rise to  a Darboux system describing an
infinite-dimensional conjugate net.
\end{abstract}

\vspace*{.5cm}

\begin{center}\begin{minipage}{12cm}

\end{minipage}
\end{center}
\newpage

\section{Introduction}
The Kadomtsev-Petviashvili(KP) equation, introduced to describe
propagation of shallow water waves under special conditions
\cite{1}, is nowadays a principal ingredient in studies of many
problems in physics and mathematics. The KP hierarchy is a paradigm
of the hierarchies of integrable systems \cite{2}-\cite{3}, it is
used to characterize Jacobian varieties in algebraic geometry
\cite{4} and is connected to string theory due to the close
relation arising between partition functions of quantum models and
KP $\tau$-functions \cite{5,7}.

The standard variables for the KP hierarchy form an
infinite-dimensional vector $\bt=(t_1,t_2,\ldots)$, the three first
components of which being the two spatial and one time variables of
the KP equation viewed as a hydrodinamical model. The KP hierarchy
can be formulated as the compatibility conditions for the following
linear system of equations (see e.g. \cite{2,3,8,9})
\begin{equation}\label{linsis}
\partial_n\psi=P_n(\bt,\partial_x)\psi,\quad \partial_n=\frac{\partial}{\partial t_n},\;
n\geq 2.
\end{equation}
Here $P_n(\bt,\partial_x)$ denotes a set of  linear differential
operators with respect to the variable $x\equiv t_1$ and
$\psi=\psi(z,\bt)$ is the KP wave function, a complex-valued
function defined on the unit circle $(|z|=1)$ which admits a
factorization $\psi=\psi_0\chi$, where
\begin{equation}\label{asex}
\psi_0=\exp(\sum_{n\geq 1}\frac{ t_n}{z^n}),\quad \chi=1+\sum_{n\geq
1}a_n(\bt)z^n.
\end{equation}

Another remarkable parametrization of the KP hierarchy is provided
by the so-called {\em Miwa variables} defined as
\begin{equation}\label{miva}
 t_n=\frac{1}{n}\sum_{i=1}^{\infty}p_i z_i^n,\quad n\geq 1,
\end{equation}
where $p_i$ and $z_i$ are discrete (integer valued) and continuous
(complex valued)  Miwa variables, respectively. It is
worth-mentioning that the  KP hierarchy  can be derived from a
single finite-difference Hirota equation involving the discrete
Miwa variables\cite{4} , which in turn is closely connected with
the Fay's trisecant formula.

Miwa variables have been extensively used several years ago in the
study of matrix models in string theory and two-dimensional quantum
gravity \cite{5}-\cite{7}. In particular, they were relevant to
show that partition functions for matrix models of Kontsevich type
are connected with KP $\tau$-functions. On the other hand, in a
different approach to string theory via the KP hierarchy it was
found that Miwa variables $p_i$ and $z_i$ can be identified with
the momenta and Koba-Nielsen variables of strings \cite{11}. At
last, it was recently shown \cite{12}that Miwa variables are also
useful in the analysis of the Bethe ansatz for quantum integrable
systems.

From the point of view of mathematics, sets of variables of the
types $t_n$ and $z_i$ connected by (\ref{miva}) have been used for
a long time in the representation theory of the symmetric group
$S_{N}$ \cite{13,14,15}. In this context $t_n$ and $z_i$ are
referred to as power sums and symmetric variables, respectively.
The presence of these variables in the theory of the KP hierarchy
is a consequence of the well established relationship between this
hierarchy and the representation theory of loop groups
\cite{15}-\cite{16}.

A main goal of the present paper is to reformulate the KP
hierarchy, its basic objects and relations, using symmetric (
continuous) Miwa variables. These variables and the spectral
parameters appear on an equal footing in the expressions for
$\tau$-functions and wave functions. Thus, it follows that several
fundamental relations for the KP hierarchy as, for instance, Sato's
equations, the relationship between wave functions and
$\tau$-functions and the action of Darboux-B\"{a}cklund transformations
take a symmetric and simple form. Furthermore,  Miwa variables are
very useful to deal with the bilinear identity for
$\tau$-functions. In particular, we prove that they provide a
compact determinantal form for addition formulae (Fay's
identities).  We also prove that the simplest Fay's identity,
written in terms of Miwa variables, leads to a Darboux system
describing an infinite-dimensional conjugate net.

\section{KP flows and $\tau$-functions}

The KP hierarchy of integrable systems can be introduced by means
of the Sato's flows \cite{sato,8,4}
\begin{equation}\label{ssato}
\displaystyle\frac{\partial S}{\partial t_n}=-(S\partial_x^n
S^{-1})_{-}S,
\end{equation}
for the pseudo-differential operator
\begin{equation}\label{ex}
S:=1+\sum_{n\geq 1}a_n(\bt)\partial_x^{-n}.
\end{equation}
The solutions of (\ref{ssato}-\ref{ex}) can be related to  KP wave
functions by setting the same coefficients $a_n(\bt)$ in the
expansions (\ref{asex}) of $\chi$ and (\ref{ex}) of $S$.

On the other hand, it is well known \cite{8},\cite{15} that a KP
wave function leads to a flow in an infinite-dimensional
Grassmannian. To describe this property we consider the Hilbert
space $H:=L^2(S^1,{\Bbb C})$ of square-integrable complex-valued
functions on the unit circle $S^1$ and the {\em big cell} $\gr_0$
of the Grassmannian $\gr$ of $H$. For our purposes $\gr_0$ can be
defined as the set of all closed subspaces $W$ of $H$ which admit a
dense linear subspace generated by an {\em admissible basis}; i.e.
a subset $\{w_n=w_n(z)\; n\geq 0\}\subset W$ of finite order
elements:
$$
 w_n(z)=z^{-n}+{\cal O}(z^{-n+1}).
$$
Each KP wave function $\psi$ determines an element $W$ of $\gr$
defined as the closure in $H$ of the set of all linear combinations
of the form
\begin{equation}\label{lc}
\sum_{n\geq 1}^N c_n(\bt_n)\psi(z,\bt_n).
\end{equation}
Here $c_n$ are arbitrary functions of $\bt$ and $\bt_n$ are
arbitrary  points of the domain of $\psi$ in ${\Bbb C}^{\infty}$ .
From (\ref{linsis}) and by using Taylor expansion around any fixed
$\bt$ it follows that
\begin{equation}\label{gr1}
 W=\mbox{span}
\{\partial_x^n\psi(z,\bt),\;n\geq 0,\; \mbox{any fixed $\bt$}\},
\end{equation}
Thus each KP wave function determines a flow in $\gr$ given by
\[
   W(\bt):=\psi_0(z,\bt)^{-1} W
=\mbox{span}\{\psi_0(z,\bt)^{-1}\partial_x^n\psi(z,\bt),\;n\geq
0,\;\}.
\]
For all value of $\bt$ in the domain of $\psi$ in ${\Bbb
C}^{\infty}$ we have that
\[
\psi_0(z,\bt)^{-1}\partial_x^n\psi(z,\bt)=z^{-n}+{\cal O}(z^{-n+1}),
\]
so that $W(\bt)\in \gr_0$.

There is a natural imbeding of $\gr $ in the projective space
$P({\cal H})$ of the infinite wedge space ${\cal
H}:=\wedge^{\infty} H$. It assigns to each $W\in \gr $ the ray
$|W\rangle$ in ${\cal H}$ generated by the vectors
$$
 w_0\wedge w_1
\wedge\cdots\wedge w_n\wedge\cdots,
$$
where $\{w_n\}_{n\geq 0}$ is any admissible basis of $W$.

 A set of fermionic operators on ${\cal H}$ can be
defined by
\[
 b_n=\frac{\partial}{\partial e_n},\quad  c_n=e_n\wedge\cdot,\;
 n=0,\pm 1,\pm 2,\ldots,
\]
where $e_n:=z_{-n}$, the operator $\frac{\partial}{\partial e_n}$
denotes the supersymmetric derivative with respect to $e_n$ and the
operator $e_n\wedge\cdot$ is the exterior multiplication operator
by $e_n$. They satisfy the canonical anticommutation relations
\[
\{b_n,b_m\}=\{c_n,c_m\}=0,\quad \{b_n,c_m\}=\delta_{nm}.
\]
Moreover, if we define the vacuum state as
\[
|vac\rangle:=|0\rangle=e_0\wedge e_1\wedge e_2\wedge\ldots\wedge
e_n\wedge
\ldots,
\]
it follows that
\begin{equation}\label{acva}
 b_n|vac\rangle=0,\; (n<0);\quad  c_n|vac\rangle=0,\; (n\geq 0).
\end{equation}
These fermionic modes determine the following fields on the unit
circle
\[
b(z):=\sum_n b_n z^{-n},\quad c(z):=\sum_n c_n z^{n},\quad |z|=1.
\]

It can be shown \cite{8} that given a KP wave function
$\psi=\psi_0\chi$ we may write
\begin{equation}\label{tau}
\chi(z,\bt)=\displaystyle\frac{\tau(\bt-[z])}{\tau(\bt)},\;\;
[z]:=(z,\frac{z^2}{2},\frac{z^3}{3},\ldots).
\end{equation}
Here the function $\tau$ is defined up to a multiplicative constant
by the correlation function
\begin{equation}\label{tau2}
\tau(\bt):=\langle vac|W(\bt)\rangle,
\end{equation}
where $W(\bt)$ is the KP trajectory of the element $W\in Gr_0$
generated by $\psi$. Moreover, it turns out \cite{8} that
\[
|W(\bt)\rangle=\exp(H(\bt))|W\rangle,\;\;H(\bt):=\sum_{n\geq 1}
t_n(\sum_{m=-\infty}^{\infty} b_m c_{m+n}).
\]

 The following bilinear identity constitutes the main property
of $\tau$-functions
\begin{equation}\label{biltau}
\int_{S^1}\exp\big(\sum_{n\geq 1}z^{-n}
(t_n-t'_n)\big)\tau(\bt-[z])\tau(\bt'+[z])\frac{dz}{z^2}=0.
\end{equation}
Natural symmetries of this identity are the so called
Darboux-B\"{a}cklund transformations \cite{17}
\begin{equation}\label{ad}
{\cal D}(p,\bt)\tau(\bt):=\psi_0(p,\bt)\tau(\bt-[p]),
\end{equation}
and their dual analogues
\begin{equation}\label{add}
{\cal D}^{*}(p,\bt)\tau(\bt):=\psi_0(p,\bt)^{-1}\tau(\bt+[p]).
\end{equation}

As a consequence of (\ref{biltau}) every KP wave function $\psi$
has an associated adjoint function $\psi^{*}=(\psi_0)^{-1}\chi^{*}$
with
\begin{equation}\label{tau3}
\chi^{*}(z,\bt)=\displaystyle\frac{\tau(\bt+[z])}{\tau(\bt)},
\end{equation}
which verifies the bilinear identity
\begin{equation}\label{bil}
\int_{S^1}\psi(z,\bt)\psi^{*}(z,\bt')\frac{dz}{z^2}=0.
\end{equation}

From (\ref{biltau}) it also follows that
\begin{equation}\label{bilca}
\int_{S^1}\psi(z,u,\bt)\psi(u',z,\bt')dz=0,
\end{equation}
where
$$
\psi(z,z',\bt)=\psi_0(z,\bt)\chi(z,z',\bt)\psi_0^{-1}(z',\bt)
$$
is the Cauchy-Baker function introduced in \cite{18} which is
determined by
\begin{equation}\label{cb}
\chi(z,z',\bt):=\displaystyle\frac{1}{z-z'}\displaystyle\frac{\tau(\bt-[z]+[z'])}{\tau(\bt)}.
\end{equation}

\section{Miwa variables}

We introduce Miwa variables $\bx=(x_1,x_2,x_3,\ldots)$ and
$\by=(y_1,y_2,y_3,\ldots)$  $|x_i|,|y_i|<1$ by
\begin{equation}\label{defmi}
\bt=\sum_{i=1}^{\infty}([x_i]-[y_i]),
\end{equation}
or equivalently
\[
t_n=\frac{1}{n}(\sum_{i=1}^{\infty}(x_i^n-y_i^n)),\quad n\geq 1.
\]
Our definition, slightly more general than those used in \cite{10}
or \cite{13,14,15}, is also adopted in \cite{6,18,19}.
Transformations of the type (\ref{defmi}) are useful in the theory
of symmetric polynomials \cite{13,14}. To show this relationship we
just mention that the Jacobian matrix of the transformation
(\ref{defmi}) is given by
\begin{equation}\begin{array}{l}
\displaystyle\frac{\partial x_i}{\partial t_n}=-(-1)^n\displaystyle
\frac{\phi_{n-1}(x_1,\ldots,\not\mbox{\hspace{-3pt}$x_i$},\ldots;y_1,y_2\ldots)}{\prod_{m\not=
i} (1-x_i/x_m)\prod_n(1-x_i/y_m)},\nonumber\\
\nonumber\\
\frac{\partial y_i}{\partial t_n}=(-1)^n\displaystyle
\frac{\phi_{n-1}(x_1,x_2,\ldots,;y_1,\ldots,\not\mbox{\hspace{-3pt}$y_i$},\ldots)}{\prod_{m\not=
i} (1-y_i/y_m)\prod_m(1-y_i/x_m)},\nonumber
\end{array}
\end{equation}
where $\phi_n$ are the symmetric functions
$$
\phi_n(z_1,z_2\ldots)=\sum_{i_1<i_2<\ldots<i_n} \frac{1}{z_{i_1}}
\frac{1}{z_{i_2}}\ldots\frac{1}{z_{i_n}}.
$$
These expressions can be derived by using the following standard
identity of the theory of symmetric polynomials
\[
\prod_{n\geq 1}(1-\frac{\lambda}{z_n})=\sum_{n\geq 0}(-\lambda)^n
\phi_n(z_1,z_2,\ldots),
\]
and taking into account that
\begin{equation}\label{tx}
\frac{\partial t_n}{\partial x_i}=x_i^{n-1},\;
\frac{\partial t_n}{\partial y_i}=-y_i^{n-1}.
\end{equation}

Given a function $\alpha(\bt)$ depending on the standard KP
variables we will denote by $\alpha(\bx;\by)$ the corresponding
function depending on the Miwa variables. In what follows the
following obvious  properties will be useful
\begin{itemize}
\item[(1)] The function $\alpha(\bx;\by)$ is invariant under
permutations among the components of $\bx$ or $\by$.
\item[(2)] $\alpha(0,\bx;\by)=\alpha(\bx;0,\by)=\alpha(\bx;\by)$
\item[(3)] For any given finite or infinite-dimensional vector $\boldsymbol{a}=(a_1,a_2,\ldots)$
\[
\alpha(\boldsymbol{a},\bx;\boldsymbol{a},\by)=\alpha(\bx;\by).
\]
\end{itemize}

Let us rewrite  the basic objects of the KP theory in terms of Miwa
variables. First of all we notice that (\ref{ssato}) and (\ref{tx})
imply that Sato's flows in Miwa variables can be written as
\begin{equation}\label{sato2}
\displaystyle\frac{\partial S}{\partial x_i}=-(S\displaystyle
\frac{\partial_x}{1-x_i\partial_x}
S^{-1})_{-}S,\;\;\;\;
\displaystyle\frac{\partial S}{\partial y_i}=(S\displaystyle
\frac{\partial_x}{1-y_i\partial_x}
S^{-1})_{-}S.
\end{equation}
The vacuum wave function takes the form
\[
\psi_0(z;\bx;\by)=\prod_{i\geq 1}\displaystyle\frac{1-y_i/z}{1-
x_i/z}.
\]
The undressed KP wave function and its adjoint function are given
by
\begin{equation}\label{neww}
\chi(z;\bx;\by)=\frac{\tau(\bx;z,\by)}{\tau(\bx,\by)},\quad
\chi^{*}(z;\bx;\by)=\frac{\tau(z,\bx;\by)}{\tau(\bx,\by)}.
\end{equation}
Analogously, the Cauchy-Baker function reads
\begin{equation}
\chi(z,z';\bx;\by)=\displaystyle\frac{1}{z-z'}\frac{\tau(z',\bx;
z,\by)}{\tau(\bx;\by)}.
\end{equation}
We would like to  emphasize that the spectral parameter $z$ appears
in the $\tau$-function on the equal footing as the Miwa variables.
It is a $y$-type variable in the expression for the wave function,
an $x$-type variable in the case of the adjoint wave function, and
it takes the role of both types in the case of the Cauchy-Baker
function. Thus, it is only when we introduce the wave functions we
mark one of the Miwa variables of the $\tau$-function, call it {\em
spectral parameter} and break its symmetry with the remaining Miwa
variables of the same type. As we will discuss elsewhere, this
symmetry between spectral parameters and Miwa coordinates at the
$\tau$-function level is very useful for dealing with symmetries of
the KP hierarchy of both standard and non-isospectral types in the
same manner.

Furthermore, it can be shown \cite{6,18,19} that the
$\tau$-function can be expressed as
\begin{equation}\label{forin}
\tau(\bx;\by)=D(\bx;\by) det\big(\displaystyle\frac{\langle
vac|c(x_i)b(y_j)|W\rangle}{y_j\langle vac|W\rangle}\big ),
\end{equation}
where
\[
D(\bx;\by):=\frac{\prod_{i,j} (y_i-x_j)}{
\prod_{n>m}(y_m-y_n)\prod_{n>m}(x_n-x_m)}.
\]
Therefore (\ref{forin}) allows us to write
\begin{equation}\label{forin2}
\tau(\bx;\by)=D(\bx;\by)det\big (
\displaystyle\frac{\tau(x_i;y_j)}{y_j-x_i}\big ).
\end{equation}
Thus, we see that the value of the $\tau$-function at a point
$(\bx;\by)\in{\boldsymbol{C}}^{2\infty}$ in Miwa space can be
expressed in terms of its values at two-dimensional single points
$(x_i,y_j)\in {\boldsymbol{C}}^2$. Notice also that the building
block of this expression for $\tau$ is closely related to the
Cauchy-Baker function as shows the identity
\[
\displaystyle\frac{\tau(x_i;y_j)}{y_j-x_i}=\tau({\boldsymbol{0}})
\chi(y_j,x_i;{\boldsymbol{0}}).
\]

Miwa variables are very suitable for dealing with the symmetry
operations of the KP hierarchy too. For instance, let us notice
that the Darboux-B\"{a}cklund transformations (\ref{ad}) take the
form
\begin{equation}
{\cal D}(p;\bx:\by)\tau(\bx;\by)=\prod_{n\geq 1}
\displaystyle\frac{1-y_n/p}{1-x_n/p}\tau(\bx;p,\by),
\end{equation}
and
\begin{equation}
{\cal D}^{*}(p;\bx:\by)\tau(\bx;\by)=\prod_{n\geq 1}
\displaystyle\frac{1-x_n/p}{1-y_n/p}\tau(p,\bx;\by).
\end{equation}
 By taking advantage of the factorized form of this expression
one easily proves the usual transformation properties of the wave
functions \cite{17}. For example, we have
\begin{equation}\begin{array}{l}
{\cal D}(p;\bx:\by)\psi(z;\bx;\by)=-\displaystyle\frac{z}{p}\psi(z;
\bx;p,\by),
\nonumber\\[.2cm]
{\cal
D}(p;\bx:\by)\psi^{*}(z;\bx;\by)=-\displaystyle\frac{p}{z}\psi^{*}(z;
\bx;p,\by),
\nonumber\\[.2cm]
{\cal
D}(p;\bx:\by)\psi(z,z';\bx;\by)=-\displaystyle\frac{z}{z'}\psi(z,z';
\bx;p,\by).\nonumber
\end{array}
\end{equation}

\section{Addition formulae and Darboux equations}

Let us consider the bilinear identity (\ref{biltau}), which written
in terms of Miwa variables takes the form
\begin{equation}\label{bil2}
\int_{S^1}\prod_{i\geq
1}\displaystyle\frac{(1-y_i/z)(1-x'_i/z)}{(1-y'_i/z)(1-x_i/z)}
\tau(\bx;z,\by)\tau(z,\bx';\by')\frac{dz}{z^2}=0.
\end{equation}
If we set
$\bx\rightarrow(\boldsymbol{s},\bx),\;\bx'\rightarrow(\boldsymbol{s'},\bx),\quad
\by'=\by$, where
\[
 \boldsymbol{s}=(s_1,s_2,\ldots,s_{N}),\quad
 \boldsymbol{s'}=(s'_1,s'_2,\ldots,s'_{M}),\quad N-M\geq 2,
 \]
and all the components of $\boldsymbol{s}$ and $\boldsymbol{s'}$
are assumed to be different, then the bilinear identity becomes
\begin{equation}\label{bil3}
\int_{S^1}\displaystyle\frac{\prod_{i=1}^M (1-s'_i/z)}{\prod_{i=1}^N(1-s_i/z)}
\tau(\boldsymbol{s},\bx;z,\by)\tau(z,\boldsymbol{s'},\bx;\by)\frac{dz}{z^2}=0.
\end{equation}
Thus by  calculating  the integral as the sum of residues at
$s_i,\;(i=1,\ldots,N)$ we get a set of Fay's identities
\begin{equation}\label{f}
\sum_{i=1}^N
\frac{\rho(s_i)}{\prod_{j\not=i}(s_i-s_j)}\hat{\tau}_i\tau_i=0,
\end{equation}
where
\[
\rho(z):=z^{N-M-2}\prod_{i=1}^M (z-s'_i),
\]
and
\[
\hat{\tau}_i:=\tau(s_1,\ldots,\not\mbox{\hspace{-3pt}$s_i$},\ldots,
s_N,\bx;\by),\quad \tau_i:=\tau(s_i,\boldsymbol{s'},\bx;\by).
\]
We note that these identities can be rewritten as
\begin{equation}\label{fay2}
\left|\begin{array}{ccccc}
1&s_1&\ldots&s_1^{N-2}&\rho(s_1)\tau_1\hat{\tau}_1\\
1&s_2&\ldots&s_2^{N-2}&\rho(s_2)\tau_2\hat{\tau}_2\\
\ldots&\ldots&\ldots&\ldots&\ldots\\
\ldots&\ldots&\ldots&\ldots&\ldots\\
1&s_N&\ldots&s_N^{N-2}&\rho(s_N)\tau_N\hat{\tau}_N\\
\end{array}\right|=0.
\end{equation}
The case $N=3,M=1$ is the usual Fay's identity
\begin{equation}\begin{array}{l}
\displaystyle\frac{s_1-s'_1}{(s_1-s_2)(s_1-s_3)}\tau(s_2,s_3,\bx;\by)\tau(s_1,s'_1,\bx;\by)\nonumber\\
\nonumber\\
+\displaystyle\frac{s_2-s'_1}{(s_2-s_1)(s_2-s_3)}\tau(s_1,s_3,\bx;\by)\tau(s_2,s'_1,\bx;\by)\nonumber\\
\nonumber\\
+\displaystyle\frac{s_3-s'_1}{(s_3-s_1)(s_3-s_2)}\tau(s_1,s_2,\bx;\by)\tau(s_3,s'_1,\bx;\by)=0.
\nonumber
\end{array}
\end{equation}
By setting $\by\rightarrow (s_2,\by)$ and dividing by
\[
\tau(s'_1,\bx;\by)\tau(s_3,\bx;\by),
\]
we get
\[ \begin{array}{l}
\displaystyle\frac{s'_1-s_1}{s_1-s_3}\chi(s_2,s_1;s'_1,\bx;\by)
+\displaystyle\frac{s'_1-s_2}{s_3-s_2}\chi(s_2,s_1;s_3,\bx;\by)\nonumber\\
\nonumber\\
+(s'_1-s_3)\chi(s_2,s_3;s'_1,\bx;\by)\chi(s_3,s_1;s_3,\bx;\by)=0.\nonumber
\end{array}
\]
We notice that a similar equation in terms of the usual KP
coordinates $t_n$ has been derived in \cite{20}. By taking the
limit $s'_1\longrightarrow s_3$ on this expression it follows
\[
\big(\frac{\partial}{\partial s_3}-
\displaystyle\frac{s_2-s_1}{(s_1-s_3)(s_2-s_3)}\big )
\chi(s_2,s_1;s_3,\bx;\by)=\chi(s_2,s_3;s_3,\bx;\by)\chi(s_3,s_1;s_3,\bx;\by).
\]
Notice that this means that for any three different  $i,j,k$ we
have
\[
\big(\frac{\partial}{\partial x_i}-
\displaystyle\frac{x_j-x_k}{(x_k-x_i)(x_j-x_i)}\big )
\chi(x_j,x_k;\bx;\by)=\chi(x_j,x_i;\bx;\by)\chi(x_i,x_k;\bx;\by).
\]
Thus, it follows at once that
\begin{equation}\label{rotcoe}
\beta_{ij}(\bx;\by):=\psi(x_i,x_j;\bx;\by)=\lim_{\epsilon\rightarrow 0}
\psi(x_i(1+\epsilon),x_j(1+\epsilon);\bx;\by),
\end{equation}
satisfy the system of Darboux equations for an infinite-dimensional
conjugate net \cite{22}
\begin{equation}\label{darboux}
\displaystyle\frac{\partial\beta_{jk}}{\partial
x_i}=\beta_{ji}\beta_{ik}.
\end{equation}

For example if we take $\tau\equiv 0$, then $\psi=\psi_0$ and we
get the following solution of the Darboux equations
\[
\beta_{ij}(\bx;\by)=\frac{1}{x_i-x_j}
\displaystyle\frac{\prod_{n\not=j}(1-x_n/x_j)}{\prod_{n\not=i}(1-x_n/x_i)}
\prod_{n\geq 1}\displaystyle\frac{1-y_n/x_i}{1-y_n/x_j}.
\]

Analogously, by taking into account that
$$
\tau=\tau(\bx;\by)\rightarrow
\tilde{\tau}(\bx;\by):=\tau(\by;\bx),
$$
is a symmetry of (\ref{biltau}), it can be proved that the
functions
\begin{equation}\label{rot2}
\tilde{\beta}_{ij}(\bx;\by):=-\psi(y_j,y_i;\bx;\by),
\end{equation}
satisfy the Darboux system
\begin{equation}\label{darboux2}
\displaystyle\frac{\partial\tilde{\beta}_{jk}}{\partial
y_i}=\tilde{\beta}_{ji}\tilde{\beta}_{ik}.
\end{equation}
The fact that the Darboux system is associated with the
one-component KP hierarchy has been already mentioned in \cite{20}
within a different approach.

We finally indicate how to get from the KP hierarchy in continuous
Miwa variables to the discrete KP hierarchy. To this end, it is
enough to constraint all the continuous Miwa coordinates
$(x_i,y_i), (i\geq 1)$ to take values on a fixed finite set
$\{a_1,\ldots,a_N\},(\; N\geq 2)$. In this way, a set of discrete
Miwa variables $(l_1,\ldots,\l_N)$ can be introduced by
\begin{equation}\label{disc}
\bt=\sum_{i=1}^{\infty}([x_i]-[y_i])=\sum_{n=1}^{N} l_n[a_n].
\end{equation}
Hence functions $\alpha(\bx;\by)$ become functions depending on the
discrete variables $l_i$. If we now consider (\ref{fay2}) with
$\boldsymbol{s}=(a_1,\ldots,a_N),\;\boldsymbol{s'}=0$, then we get
\begin{equation}\label{faydis}
\left|\begin{array}{ccccc}
1&a_1&\ldots&a_1^{N-2}&a_1^{N-2}\tau_1\hat{\tau}_1\\
1&a_2&\ldots&a_2^{N-2}&a_2^{N-2}\tau_2\hat{\tau}_2\\
\ldots&\ldots&\ldots&\ldots&\ldots\\
\ldots&\ldots&\ldots&\ldots&\ldots\\
1&a_N&\ldots&a_N^{N-2}&a_N^{N-2}\tau_N\hat{\tau}_N\\
\end{array}\right|=0,
\end{equation}
where
\[\begin{array}{l}
\hat{\tau}_i:=\tau(l_1+1,\ldots,l_{i-1}+1,l_i,l_{i+1}+1,\ldots,
l_N+1),\\
\tau_i:=\tau(l_1,\ldots,l_{i-1},l_i+1,l_{i+1},\ldots,l_N).
\end{array}
\]
These equations constitute the discrete KP hierarchy \cite{21}.

\end{document}